\newcommand{\bx}{{\bf x} }
\newcommand{\p}{\partial}

\newcommand{\tr}{\tilde{r}}

\newcommand{\tpsi}{\tilde{\psi}}
\newcommand{\tmu}{\tilde{\mu}}
\newcommand{\gm}{\gamma}
\newcommand{\vep}{\varepsilon}
\newcommand{\be}{\begin{equation}}
\newcommand{\ee}{\end{equation}}

\documentclass[%
 reprint,
superscriptaddress,
showpacs,preprintnumbers,
 amsmath,amssymb,
 aps,
floatfix,
]{revtex4-1}

\usepackage{color} 

\usepackage{graphicx}
\usepackage{dcolumn}
\usepackage{bm}

\usepackage[caption=false]{subfig}
\graphicspath{{./figure/}}

\begin{document}

\preprint{APS/123-QED}

\title{Mean-field regime and Thomas-Fermi approximations of trapped Bose-Einstein
condensates with higher order interactions in one  and two dimensions}

\author{Xinran Ruan}
\affiliation{Department of Mathematics, National
University of Singapore, Singapore 119076}
\author{Yongyong Cai}
\email{yongyong.cai@gmail.com}
\affiliation{Beijing Computational Science Research Center, Beijing 100094,
P. R. China}
\affiliation{Department of Mathematics, Purdue University, West Lafayette,
IN 47907, USA}
\author{Weizhu Bao}
\email{matbaowz@nus.edu.sg}
\affiliation{Department of Mathematics, National
University of Singapore, Singapore 119076}

\date{\today}
\begin{abstract}
We derive rigorously one- and two-dimensional mean-field equations for
  cigar- and pancake-shaped  Bose-Einstein condensates (BEC) with higher order interactions (HOI).
 We show how the  higher order interaction modifies the contact interaction of
  the strongly confined particles. Surprisingly, we find that the usual Gaussian profile assumption for the strongly confining direction
  is inappropriate for
  the cigar-shaped BEC case, and a Thomas-Fermi type profile should be adopted instead.
  Based on the derived mean field equations, the Thomas-Fermi densities are analyzed in presence of the contact interaction and HOI.
  For both box and harmonic traps in one, two and three dimensions, we identify the analytical Thomas-Fermi densities, which depend on the competition between
  the  contact interaction and the  HOI.
\end{abstract}

\maketitle


\section{\label{sec:level1}Introduction}
  Quantum-degenerate gases have been extensively explored since the
  remarkable realizations of Bose-Einstein Condensate (BEC) in 1995 \cite{Anderson,Davis,Bradley}.
  In the typical experiments of BEC, the ultra-cold bosonic  gases
  are dilute and weakly interacting, and yet the major properties of the system are governed
  by these weak two-body  interactions \cite{Dalfovo,Pethick,Pitaevskii}. Though the atomic interaction
  potentials are rather complicated, they can be effectively described by
 the two-body Fermi contact interaction -- interaction kernel taken as the
 Dirac delta  function -- in the ultra cold dilute regime,
 with a single parameter, the zero energy $s$-wave scattering
 length $a_s$. This is the heart of the mean field Gross-Pitaevskii equation (GPE)
 theory for BEC \cite{Pitaevskii,Lie}.  Based on GPE, various aspects of
 BEC have been extensively studied, including the static
 properties \cite{Lie,Hau,Bao2013} and dynamical properties \cite{Holl,Stamper,BaoJakschP}.

 The treatment of effective two-body contact interactions has been proven to be successful,
 but it is limited due to the low energy  or low density assumption \cite{Esry}.
In the case of  high particle densities or strong confinement,   there will be a wider
range of possible momentum states  and  correction terms should be
included in the GPE for better description \cite{Fu,An}. Within the
perturbative framework,
higher order interaction (HOI) (or effective range expansion) as a
correction to the delta function,  has to be taken into account, resulting
in a modified Gross-Pitaveskii equation (MGPE), e.g. Eq. \eqref{mgpe} \cite{Fu,Collin,Veksler}.

Based on  the MGPE \eqref{mgpe}, \cite{Qi,Qix,Wamba}
have shown the stability conditions and collective excitations of a
harmonically trapped BEC. In the Thomas-Fermi(TF) limit regime,   \cite{Tho} has shown the
 approximate density profile for BEC with HOI in a radial trap. On the other hand,
in most experiments,  a strong harmonic trap along one or two directions confine
(or suppress) the condensate  into pancake or cigar shape, respectively.
In such cases, the usual TF approximation for the full three-dimensional (3D) case
becomes invalid. It is then desirable to  derive the effective one- (1D) and two-dimensional (2D) models,
which offers compelling advantage for numerical computations compared to the 3D case.

In this paper,  we present effective mean-field equations for trapped
BECs with HOI in one and two dimensions.  Our equations are based on
a mathematically rigorous dimension reduction of the 3D
 MGPE \eqref{mgpe} to lower dimensions. Such dimension reduction has been formally derived in \cite{Bao2007,Ge,Sala,YSA,CRLB,Bao2013} and rigorously
 analyzed in \cite{Ben,BLM}, for the conventional GPE, i.e. without HOI.
 While for the MGPE, to our knowledge, this result has not been obtained, except for
 some initial work \cite{Wamba,Veksler}, where the Gaussian profile is assumed in the strongly confining direction following the conventional GPE case.
Surprisingly, our findings suggest that the Gaussian profile assumption is inappropriate for the quasi-1D BEC.
 In the derivation of
the quasi-1D (2D) model for the BEC with HOI, we assume that the leading order (in terms of aspect ratio)
of the full 3D energy is from the radial (longitudinal) wave function,
such that the BEC can only be excited in the  non-confining directions,
resulting in effective 1D (2D)
condensates. Based on this principle, we show that the longitudinal wave function
can be taken as the ground state of the longitudinal harmonic trap in quasi-2D BEC, and the radial
wave function has to be taken as the Thomas-Fermi (TF) type (see \eqref{3to1factor})  in
quasi-1D BEC, which is totally different from the conventional GPE case  \cite{Wamba,Veksler}.  Furthermore,
we derive simple TF densities in 1D, 2D and 3D from our effective
equations, with different HOI  and contact interaction parameters, for both harmonic potentials and box potentials.
These results demonstrate very interesting phase diagrams of the TF ground state densities regarding the contact interaction and HOI.
We compare the ground states of the quasi-1D and quasi-2D
BEC with the ground states of the full 3D BEC and find good
agreement.  In particular, our ground states are  good approximations
to those of the full 3D MGPE in regimes where the TF
approximation fails.

 The paper is organized as follows. In Sec.~\ref{sec:model}, we introduce the modified GPE in presence of
HOI that will be considered in this paper.
 As the first main
result, we present  in Sec.~\ref{sec:3to1} a mean-field equation for a
quasi-1D cigar-shaped BEC.  We compare the ground state solutions of this
1D equation with the full 3D computation.
 In Sec.~\ref{sec:3to2}, we present the second main result, a mean-field equation for a quasi-2D   pancake-shaped BEC. Comparisons are
 made between ground state density profiles of the 2D equation and ground state density profiles calculated  from the full 3D model.
 In Sec.~\ref{sec:har}, we provide a complete summary of the TF approximation in 1D, 2D and 3D cases,   with harmonic potential or box potential separately.
Depending on  the HOI strength  and contact interaction strength, TF approximate solutions are surprisingly different, which are compared with the corresponding
 ground state solutions obtained via mean field equations. Finally, we present our conclusions in Sec.~\ref{sec:conclusion}. Appendix ~\ref{appendix:3to1} provides the
 details of the dimension reduction from full 3D MGPE to our 1D mean-field equation, and Appendix ~\ref{appendix:3to2} provides the details for the reduction to 2D case.

\section{\label{sec:model}3D modified GPE}
At the temperature $T$ much smaller than the critical temperature $T_c$, the mean-field Gross-Pitaevskii equation (GPE) of a BEC
can be combined with the HOI effect \cite{Fu,Wamba}. Inserting the HOI corrections to the two-body interaction potential,
we can obtain  the modified Gross-Pitaevskii equation (MGPE) \cite{Collin,Veksler} for the wave function $\psi:=\psi(\bx,t)$ as
\begin{equation}\label{mgpe}
i\hbar\partial_t\psi=\left[-\frac{\hbar^2}{2m}\nabla^2+V+g_{0}(|\psi|^2+g_1\nabla^2|\psi|^2)\right]\psi,
\end{equation}
where $\bx=(x,y,z)^T\in \mathbb R^3$   is the Cartesian coordinate
vector, $\hbar$ is the reduced Planck constant, $m$ is the mass of the particle,
$g_0=\frac{4\pi \hbar^2 a_{s}}{m}$ is the
contact interaction strength with $a_s$ being the $s$-wave scattering
lengths, HOI correction is given by the parameter
$g_1=\frac{a_s^2}{3}-\frac{a_sr_e}{2}$ with $r_e$ being the effective range of the two-body interactions and
$r_e=\frac{2}{3}a_s$ for hard sphere potential,
$V:=V(\bx)$ is the given
real-valued external trapping potential.
As in typical current experiments, we assume BEC is confined in the following harmonic potential
\begin{equation}\label{trap}
V(\bx)=\frac{m}{2}\left[\omega_x^2x^2+\omega_y^2y^2+\omega_z^2z^2\right],
\end{equation}
where $\omega_x>0$, $\omega_y>0$ and $\omega_z>0$ are trapping frequencies in $x$-, $y$- and $z$-direction,
respectively.
The wave function $\psi$ is  normalized as
\begin{equation} \label{norm}
\|\psi(\cdot,t)\|^2:=\int_{{\mathbb R}^3}|\psi(\bx,t)|^2d\bx=N,
\end{equation}
where $N$ is the total number of particles in BEC.

We introduce the dimensionless quantities by rescaling  length,  time, energy and
wave function as
$\bx\to \bx x_s$, $t\to t/\omega_0$, $E\to E \hbar\omega_0$ and $\psi\to\psi\sqrt{N/x_s^3}$, respectively,
where $x_s=\sqrt{\frac{\hbar}{m\omega_0}}$ with $\omega_0=\min\{\omega_x,\omega_y,\omega_z\}$, $E$ is the energy.
After rescaling, the dimensionless form of the MGPE \eqref{mgpe} reads
\begin{equation}\label{eq:mgpe}
i\partial_t\psi=-\frac{1}{2}\nabla^2\psi+V(\bx)\psi+\beta|\psi|^2\psi-\delta\nabla^2(|\psi|^2)\psi,
\end{equation}
with
\begin{equation}
\quad\beta=4\pi{N}\frac{a_s}{x_s},\quad\delta=-\frac{4\pi{N}}{x_s^3}\left(\frac{a_s^3}{3}-\frac{a_s^2r_e}{2}\right),
\end{equation}
and the dimensionless trapping potential is $V(\bx)=\gm_x^2x^2/2+\gm_y^2y^2/2+\gm_z^2z^2/2$
with $\gm_x=\omega_x/\omega_0,\quad\gm_y=\omega_y/\omega_0,\quad\gm_z=\omega_z/\omega_0$. The normalization condition becomes
\be\label{eq:normalization}
\|\psi(\cdot,t)\|^2=\int_{\mathbb{R}^3}|\psi(\bx,t)|^2d\bx=1.
\ee

When $\delta=0$, the MGPE \eqref{eq:mgpe} collapses to the  conventional GPE
and the corresponding dimension reduction problem has been studied in
\cite{Bao2013,Bao2007,Ge,Sala,YSA,CRLB,Ben} and references therein.
When $\delta<0$, there is no ground state of \eqref{eq:mgpe},
and when $\delta>0$ and the trapping potential is a confinement,
there exist ground states of \eqref{eq:mgpe} and the positive ground state is unique if $\beta\ge0$.
Thus hereafter we assume $\delta>0$.

\section{\label{sec:3to1}quasi-1D BEC with HOI}
With a sufficiently large radial trapping frequency, it is possible to freeze the radial motion of BEC \cite{Gorlitz}, which becomes a quasi-1D system. Intuitively, the  energy separation between stationary states is  much larger in the radial direction than in the axial direction, and the dynamics is then freezed in radial direction.
 As a consequence,  the wave function of the system is in the variable separated form, i.e., it is  the multiplication of the  axial direction function and the   radial direction function.  In
this section, we present an effective mean-field equation for the
axial wave function of such a strongly confined BEC with HOI, by assuming a strong radial confinement.

\subsection{1D mean-field equation}
In order to derive the mean-field equation for the axial wave function, we start with the 3D MGPE \eqref{mgpe} and assume a harmonic potential with
$\omega_r=\omega_x=\omega_y\gg\omega_z$. Choosing rescaling parameters used in \eqref{eq:mgpe}  as $\omega_0=\omega_z$, $x_s=\sqrt{\hbar/m\omega_z}$, we now work
with the dimensionless equation \eqref{eq:mgpe}.
In the quasi-1D BEC with HOI, the 3D wavefunction can be factorized as
\begin{equation}\label{3to1factor}
\psi(\bx,t)=e^{-i\mu_{2D}t}\omega_{2D}(x,y)\psi_{1D}(z,t),
\end{equation}
with appropriate radial state function $\omega_{2D}$ and $\mu_{2D}\in\Bbb R$. Once  the radial state $\omega_{2D}$ is known, we could project the MGPE \eqref{eq:mgpe}
onto the axial direction to derive the quasi-1D equation. The key to find such $\omega_{2D}$ is the criteria that, the energy separation between stationary states should be much larger in the radial direction than in the axial direction, i.e., there is energy  scale separation between the radial state $\omega_{2D}$ and axial wavefunction.

We denote  the aspect ratio of the harmonic trap as
\begin{equation}
\gamma=\omega_r/\omega_z.
\end{equation}
For conventional GPE, i.e., $\delta=0$,  a good choice for $\omega_{2D}$ is the Gaussian function \cite{Bao2013}, which is the ground state of the radial harmonic trap,  as $\omega_{2D}(r)=\sqrt{\frac{\gm}{\pi}}e^{-\frac{\gm{r}^2}{2}}$.  The reason  is that the order of energy separation between states of conventional BEC   is dominated in the  radial direction by the radial harmonic oscillator part, which is $O(\gm)$, much larger than the interaction energy part if $\beta=O(1)$ by a similar computation shown in Appendix \ref{appendix:3to1}. Alternatively,  it would be possible to use variational Gaussian profile approach to find $\omega_{2D}(r)$ \cite{Sala}.  However, for the BEC with HOI case, the extra HOI  term contributes to the energy. Thus, a more careful comparison between the kinetic energy part and the HOI  energy part  is demanded.

By a detailed computation (see Appendix \ref{appendix:3to1}),  we identify the energy contribution from the HOI term \eqref{rmodel} in transverse direction,  is dominant as $\gamma\gg1$. It shows a completely different scenario  compared to  the conventional GPE, in which the transverse harmonic oscillator terms are dominant. The explicit form for the  transverse radial state function $\omega_{2D}(r)$ for quasi-1D BEC with HOI is  determined as
\begin{equation}
\omega_{2D}(x,y)\approx\frac{\gm(R^2-r^2)_+}{4\sqrt{2\delta_{r}}},\quad r=\sqrt{x^2+y^2},\label{eq:dt2}
\end{equation}
where  $R=2\left(\frac{3\delta_r}{2\pi\gm^2}\right)^{\frac{1}{6}}$,
$\delta_{r}=\frac{2\cdot 3^{\frac{5}{7}}\pi^{\frac{1}{7}}\delta^{\frac{6}{7}}}{5^{\frac{9}{7}}\gm^{\frac{4}{7}}}$, $\mu_{2D}{\approx\frac{3^{\frac{4}{7}}\delta^{\frac{2}{7}}\gm^{\frac{8}{7}}}{\pi^{\frac{2}{7}}5^{\frac{3}{7}}}}$
 and $(f)_+=\max\{f,0\}$.

It is worth pointing out  that the  determination of the radial state $\omega_{2D}(r)$
is coupled with the axial direction state (see (\ref{para:bd})). Therefore, a coupled system of the radial and axial states is necessary  to get  refined approximate density profiles for ground states, instead of using the above approximate $\omega_{2D}(r)$.

In the axial $z$ direction,  multiplying \eqref{eq:mgpe} by $\omega_{2D}$ and integrating the $x,y$ variables out,   we obtain the mean-field equation for quasi-1D BEC with HOI as
\begin{widetext}
\begin{equation}\label{3to1model}
i\partial_t\psi_{1D}(z,t)=-\frac{1}{2}\partial_{zz}\psi_{1D}+V_{1D}(z)\psi_{1D}
+\beta_1|\psi_{1D}|^2\psi_{1D}-\delta_1(\partial_{zz}|\psi_{1D}|^2)\psi_{1D},
\end{equation}
\end{widetext}
where $V_{1D}(z)=\frac12\gamma_z^2z^2=\frac12z^2$, and
\begin{subequations}\label{eq:1dpara}
\begin{align}
\beta_1&=\frac{5^{\frac{6}{7}}}{3^{\frac{1}{7}}\cdot4\pi^{\frac{3}{7}}}\delta^{\frac{3}{7}}\gm^{\frac{12}{7}}+
\frac{3^{\frac{10}{7}}}{4\cdot5^{\frac{4}{7}}\pi^{\frac{5}{7}}}
\frac{\beta\gm^{\frac{6}{7}}}{\delta^{\frac{2}{7}}},\label{eq:beta1}\\
\delta_1&=\frac{3^{\frac{10}{7}}}{4\cdot5^{\frac{4}{7}}\pi^{\frac{5}{7}}}\delta^{\frac{5}{7}}\gm^{\frac{6}{7}}.
\end{align}
\end{subequations}
From Eq. \eqref{3to1model}, it is observed that the HOI provides extra repulsive contact interactions in the quasi-1D BEC. More interestingly, the first term in $\beta_1$ suggests that
the contact interactions is dominated by HOI part.

  If the repulsive contact interaction dominates the dynamics in \eqref{3to1model}, we could neglect the kinetic and HOI parts to obtain an analytical expression for
  the quasi-1D BEC with HOI. This agrees with the usual Thomas-Fermi approximation for conventional quasi-1D BEC, and its validity is  shown in Sec.~\ref{sec:har} (referred as region I). In such situation, the approximate density profile is given as:
\begin{equation}\label{eq:denstiy1d}
n_{1D}(z)=|\psi_{1D}|^2=\frac{\left((z^{*})^2-z^2\right)_+}{2\beta_1},
\end{equation}
where $z^*=\left(\frac{3\beta_1}{2}\right)^{\frac{1}{3}}$.


\begin{figure}
\includegraphics[width=.48\textwidth]{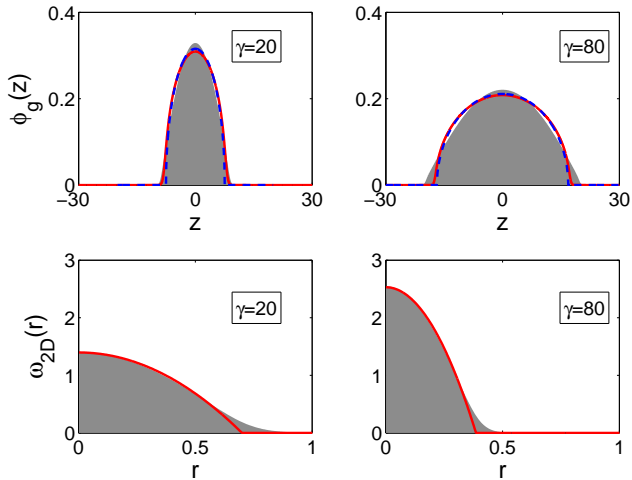}
\caption{(quasi-1D ground state) Red line: approximation \eqref{eq:dt2}	in radical direction and numerical solution of \eqref{3to1model} in axial direction. Blue dash line: Thomas-Fermi approximation of \eqref{eq:denstiy1d} in axial direction. Shaded area: numerical solution from the original 3D model \eqref{eq:mgpe}.
The corresponding $\gm$'s are given in the plots. For other parameters, we choose $\beta=1,\delta=20$.}
\label{fig:cigar}
\end{figure}




In Fig.~\ref{fig:cigar}, we compare the ground state densities of quasi-1D BEC with HOI determined via \eqref{3to1model}, analytical predication \eqref{eq:denstiy1d} and the numerical results from 3D MGPE in \eqref{eq:mgpe} by integrating over the transversal directions. Noticing that HOI term produces effective repulsive potential,  the BEC is broadened  compared to the analytically predicated profile. As a consequence, in the regime of small or moderate interaction energy $\beta_{1}$,
  we predict that the usual approach to  BECs with HOI via
conventional TF approximation fails. On the other hand, our proposed 1D equation, Eq. \eqref{3to1model}, describes the BEC accurately in the mean-
field regime at experimentally relevant trap aspect ratios $\gamma$.


\section{\label{sec:3to2}quasi-2D BEC with HOI}
In this section, we consider the BEC being strongly confined in $z$ axis, which corresponds to $0<\gm\ll1$. Accordingly, we choose rescaling parameters used in \eqref{eq:mgpe}  as $\omega_0=\omega_r$, $x_s=\sqrt{\hbar/m\omega_r}$, and we  work with the dimensionless equation \eqref{eq:mgpe}.

 Similarly to the case of quasi-1D BEC, we assume that  the   wave function can be factorized in the quasi-2D case, as
\begin{equation}\label{3to2factor}
\psi(\bx,t)=e^{-i\mu_{1D}t}\psi_{2D}(x,y,t)\omega_{1D}(z),
\end{equation}
for appropriate longitudinal state $\omega_{1D}(z)$ and $\mu_{1D}\in\mathbb{R}$.

Following the same procedure as that for the quasi-1D BEC case, we find  that,   the leading order energy separations  in $z$ direction is  due to the longitudinal harmonic oscillator, while the cubic interaction and HOI parts are less important (see Appendix \ref{appendix:3to2} for details). This fact suggests that the ground mode of the longitudinal harmonic oscillator is a suitable choice for $\omega_{1D}(z)$, i.e., a Gaussian type function as
\begin{equation}\label{eq:w1d}
\omega_{1D}(z)\approx\left(\frac{1}{\pi\gm}\right)^{\frac{1}{4}}e^{-\frac{z^2}{2\gm}},
\end{equation}
and $\mu_{1D}\approx 1/2\gamma$.


Substituting \eqref{3to2factor} with \eqref{eq:w1d} into the MGPE \eqref{eq:mgpe}, then multiplying \eqref{eq:mgpe} by $\omega_{1D}$ and integrating the longitudinal $z$ out, we obtain a mean-field equation for quasi-2D BEC with HOI  as
\begin{widetext}
\begin{equation}\label{3to2model}
i\partial_{t}\psi_{2D}=-\frac{1}{2}\nabla^2\psi_{2D}+V_{2D}(x,y)\psi_{2D}+\beta_2|\psi_{2D}|^2\psi_{2D}-\delta_2(\nabla^2|\psi_{2D}|^2)\psi_{2D},
\end{equation}
\end{widetext}
where $V_{2D}(x,y)=\frac{1}{2}(x^2+y^2)$ and
\begin{align}
\beta_2&=\frac{\beta}{\sqrt{2\pi\gm}}+\frac{\delta}{\sqrt{2\pi\gm^3}},
\quad\delta_2=\frac{\delta}{\sqrt{2\pi\gm}}.\label{beta_2}
\end{align}
Similarly to the quasi-1D BEC case, HOI induces effective contact interactions in the quasi-2D regime, which dominates the contact interaction ($\beta$ part). We then conclude that
even for small HOI $\delta$, the contribution of HOI could be significant in the high particle density regime of quasi-2D BEC.

Analogous to the quasi-1D BEC case,  we can derive the usual Thomas-Fermi (TF) approximation when the repulsive interaction $\beta_2$ dominates the dynamics, and
 the analytical densities for the quasi-2D BEC with HOI reads
\begin{equation}\label{eq:tf2d}
n_{2D}(r)=|\psi_{2D}|^2=\frac{\left(R^2-r^2\right)_+}{2\beta_2},\quad r=\sqrt{x^2+y^2},
\end{equation}
 where $R=\left(\frac{4\beta_2}{\pi}\right)^{\frac{1}{4}}$.

%

\begin{figure}
\includegraphics[width=.48\textwidth]{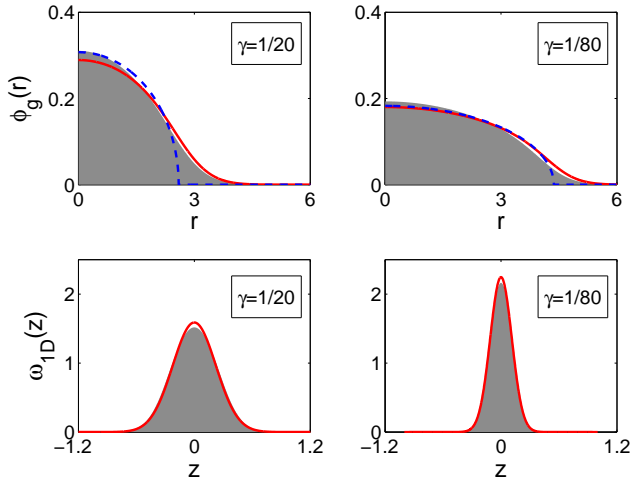}
\caption{(quasi-2D ground state) Red line:  approximation \eqref{eq:w1d} in axial direction and numerical solution of \eqref{3to2model} in radical direction. Blue dash line: Thomas-Fermi approximation of \eqref{eq:tf2d} in radical direction. Shaded area: numerical solution from the original 3D model \eqref{eq:mgpe}.
The corresponding $\gm$'s are given in the plots. For other parameters, we choose $\beta=5,\delta=1$.}
\label{fig:disk}
\end{figure}

In order to verify our findings in this section, we compare the quasi-2D ground state densities obtained via Eq. \eqref{3to2model}, TF density \eqref{eq:tf2d} and
the numerical results from 3D MGPE \eqref{eq:mgpe} by integrating the longitudinal $z$ axis out. The results are displayed in Fig.~\ref{fig:disk}. Similarly to the quasi-1D case,   the BEC is broadened  compared to the analytically predicated profile because of the effective repulsive interaction from the HOI. Thus, in the regime of small or moderate interaction energy $\beta_{2}$,
 the usual approach to  BECs with HOI via conventional Thomas-Fermi approximation fails. On the other hand,  it turns out that our proposed 2D equation, Eq. \eqref{3to2model}, is accurate for quasi-2D BEC in the mean-field regime at experimentally relevant trap aspect ratios $\gamma$.

\section{\label{sec:har}Thomas-Fermi (TF) approximation}
In the previous sections, we have derived 1D \eqref{3to1model} and 2D \eqref{3to2model} equations for the quasi-1D and quasi-2D BECs, respectively. Indeed, all the 1D \eqref{3to1model}, 2D \eqref{3to2model} and 3D \eqref{eq:mgpe} equations can be written in a unified form as
\begin{equation}\label{eq:mgpe:d}
i\partial_t\psi=-\frac{1}{2}\nabla^2\psi+V(\bx)\psi+\beta|\psi|^2\psi-\delta\nabla^2(|\psi|^2)\psi,
\end{equation}
where $\bx\in\mathbb{R}^d$, $d=3, 2, 1$, $\beta$ and $\delta$ are treated as parameters ($\delta$ is positive). Though $V(\bx)$ is assumed to be harmonic potential in the previous derivation, it is not necessary to restrict ourselves for the harmonic potential case. Thus, we treat $V(\bx)$ as a general real-valued potential in this section. In particular, we will address the cases when $V(\bx)$
is a radially symmetric harmonic potential as
\begin{equation}
V(\bx)=\frac12\gamma_{0}^2r^2, \qquad r=|\bx|,
\end{equation}
where $\gamma_0>0$ is a dimensionless constant,
or a radial box potential as
\begin{equation}
V_{\rm box}(\bx)=\begin{cases}0,&0\le r< R,\\
\infty, &r\ge R.
\end{cases}
\end{equation}

As pointed out in the quasi-1D, 2D cases, a dominant repulsive contact interaction will lead to an analytical  TF densities, analogous to the conventional BEC case. However, with HOI \eqref{eq:mgpe:d}, the system is characterized by two interactions, contact interaction strength $\beta$ and HOI strength
$\delta$, which is totally different from the classical GPE theory that the BEC is purely characterized by the contact interaction $\beta$. Hence, for BEC with HOI \eqref{eq:mgpe:d}, it is possible that HOI interaction competes with contact interaction, and may be the
major effect determining the properties of BEC. In this section, we will discuss how the competition between $\beta$ and $\delta$ leads to different density profiles for the strong interactions, for which we refer such analytical density approximations as the TF approximations. We notice that it might not be physical to consider HOI as the key factor  of BEC
in three dimensions in  current BEC experiments, but we  treat $\delta$ and $\beta$ in \eqref{eq:mgpe:d}  as arbitrary parameters and the result presented here may find its application in future and/or in the other fields.

In previous sections on quasi-1D and 2D BECs, we have given the analytical TF densities for $\beta$ dominant system.  For the general consideration of the large $\beta$ and $\delta$
interactions, we show in Fig.~\ref{fig:regime} the phase diagram of the different parameter regimes for $\beta$ and $\delta$, in which the TF approximation are totally different. Intuitively, there are three of them: $\beta$ term is more important (regime I in Fig.~\ref{fig:regime}), $\delta$ term is more important (regime III), and $\beta$ term is comparable to the $\delta$ term (regimes II \& IV).  Detailed computations and arguments for the results shown in Fig.~\ref{fig:regime} can be found in the Appendix~\ref{appendix:TF}. Based on Fig.~\ref{fig:regime}, we will discuss the harmonic potential and the box potential cases separately.

\subsection{TF approximation with harmonic potential}
From Fig.~\ref{fig:regime}(a),  the curve $\beta=O(\delta^{\frac{d+2}{d+4}})$ is the boundary that divides the regimes for harmonic potential case. To be more specific, if $\beta\gg\delta^{\frac{d+2}{d+4}}$, the cubic nonlinear term is more important, and vise versa. If $\beta= O(\delta^{\frac{d+2}{d+4}})$, both of the two nonlinear terms are important, and have to be taken care of in the TF approximation. The resulting analytical TF density profiles in different regimes, are listed below:
\begin{figure}
\centering
\begin{tabular}{c}
\subfloat{\includegraphics[width=.48\textwidth]{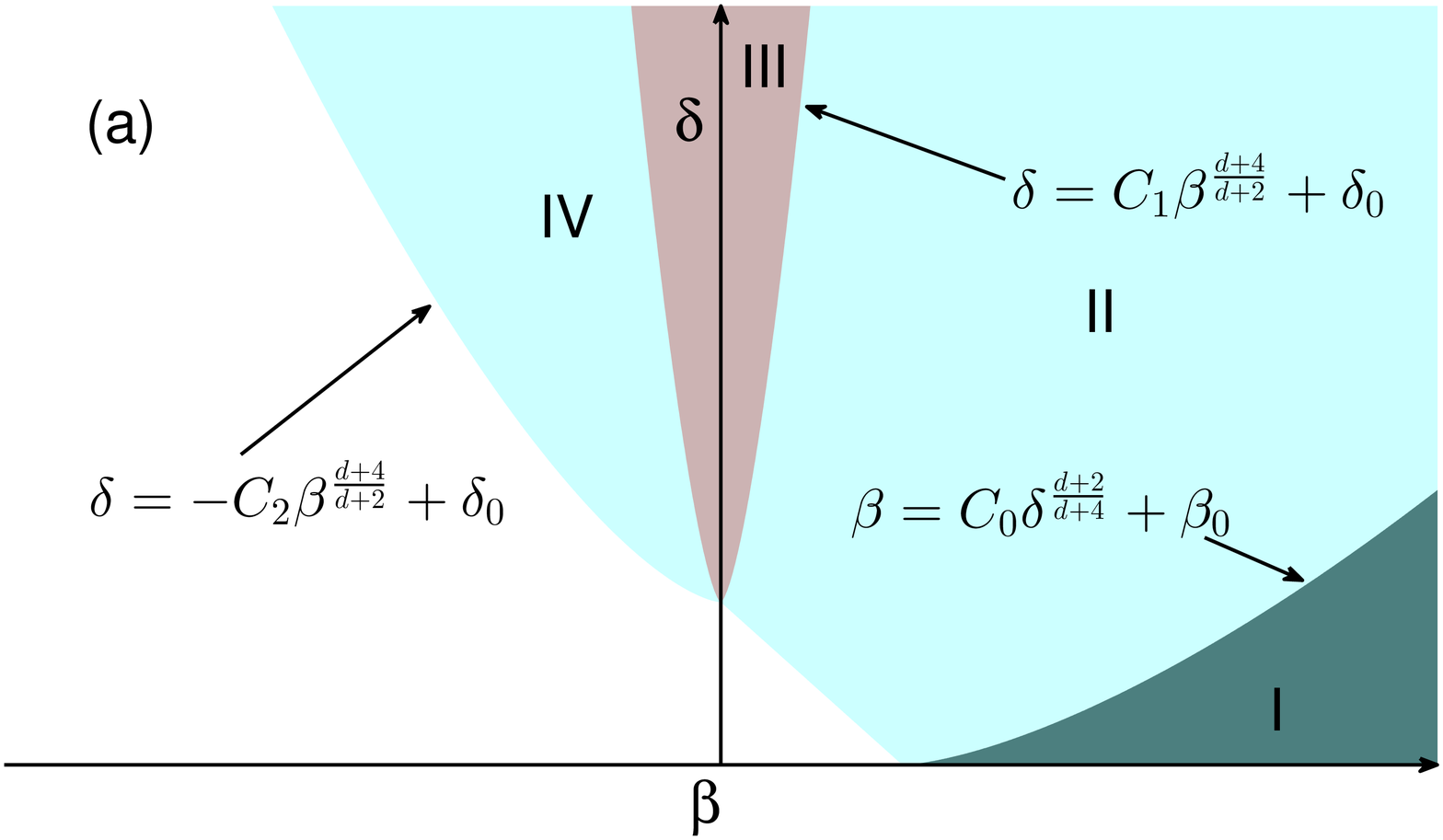}}\\
\subfloat{\includegraphics[width=.48\textwidth]{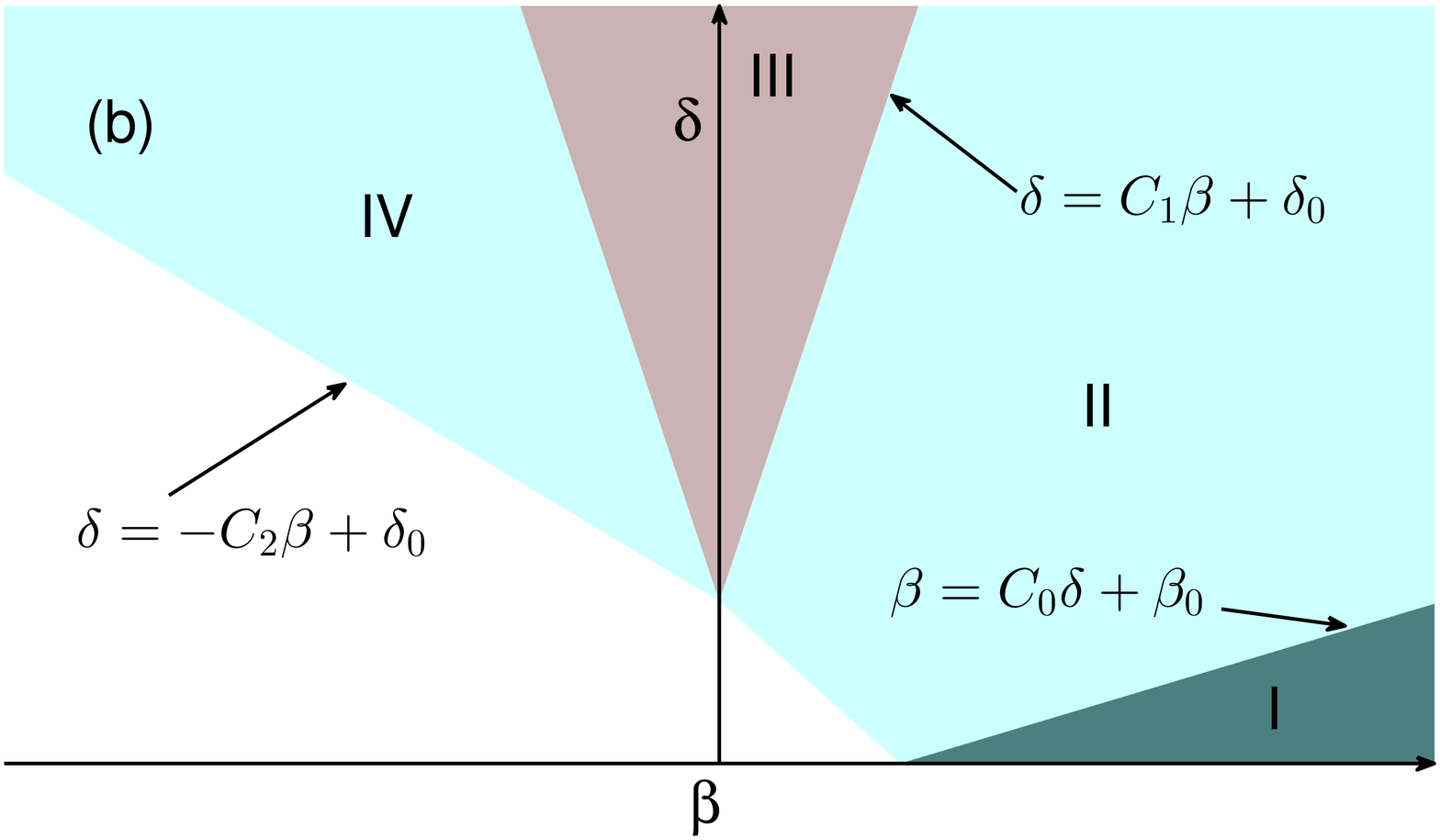}}
\end{tabular}
\caption{Phase diagram for extreme regimes: (a) is for harmonic potential case and (b) is for box potential case.
In the figure, we choose $\beta_0\gg1$ and $\delta_0\gg1$.}
\label{fig:regime}
\end{figure}

{\bf Regime I}, i.e. $\beta\gg\delta^{\frac{d+2}{d+4}}$, the $\delta$ term and the kinetic energy term are dropped, and the density profile is determined as
\begin{equation}\label{ground_sol_har_lb}
n_{\rm TF}(r)=|\psi_{\rm TF}|^2=\frac{\gm_0^2(R^2-r^2)_+}{2\beta},
\end{equation}
where $R=\left(\frac{(d+2)C_d\beta}{\gm_0^2}\right)^{\frac{1}{d+2}}$, and  the constant $C_d$ is defined as
\begin{equation}\label{eq:cd}
C_d=
\begin{cases}
\frac{1}{2},& d=1,\\
\frac{1}{\pi},& d=2,\\
\frac{3}{4\pi},& d=3.
\end{cases}
\end{equation}
With the above TF densities, the leading order approximations for chemical potential $\mu$ and energy $E$ of the ground state are: $\mu_{\rm TF}=\frac{1}{2}\left((d+2)C_d\beta\right)^{\frac{2}{d+2}}\gm_0^{\frac{2d}{d+2}}$, $E_{\rm TF}=\frac{d+2}{d+4}\mu_{\rm TF}$
for  $d$ ($d=3,2,1$) dimensional case.

{\bf Regime II}, i.e. $\beta=C_0\delta^{\frac{d+2}{d+4}}$ with $C_0>0$, neglecting the kinetic  term in the time-independent MGPE, we have
\begin{equation}\label{eq:har_bd}
\mu\psi=\frac{\gm_0^2|\bx|^2}{2}\psi+C_0\delta^{\frac{d+2}{d+4}}|\psi|^2\psi-\delta\nabla^2(|\psi|^2)\psi.
\end{equation}
Formally, Eq. \eqref{eq:har_bd} degenerates at position $\bx$ if $\psi(\bx)=0$ and it is indeed a free boundary problem (boundary of the zero level set of $\psi$), which
  requires careful consideration. Motivated by \cite{Tho} for the 3D case,
besides the condition that $n(R)=0$ along the free boundary  $|\bx|=R$, we impose $n^{\prime}(R)=0$;
and assume $n(r)=0$ for $r>R$.

The TF density profile in regime II  is self similar under appropriate scalings.
To be more specific, the analytical TF density takes the form
\be\label{eq:TFform}
n_{\rm TF}(r)=|\psi_{\rm TF}|^2=\delta^{-\frac{d}{d+4}}n_{0}(\delta^{-\frac{1}{d+4}}r),
\ee
 where $n_0(r)$ is the function can be calculated exactly as below.

 Plugging \eqref{eq:TFform} into \eqref{eq:har_bd}, we obtain the equation for $n_0(r)$ by imposing the aforementioned conditions at the free boundary,
\be\label{eq:har_bd_new}
\tmu=\frac{\gm_0^2r^2}{2}+{C_0}n_0-\partial_{rr}n_0(r)-\frac{d-1}{r}\partial_r n_0(r),
\ee
for $r\leq R$ and $n_0(s)=0$ for $s\ge R$, and $n_0(R)=0$, $n_0^\prime(R)=0$, where $R$ is the free boundary that has to be determined and $\tmu=\delta^{-\frac{2}{d+4}}\mu$.
In addition, we assign the boundary condition
at $r=0$ as $n_0^\prime(0)=0$, because of the symmetry.

Note that $C_0$ can be negative as $\delta$ term can bound the negative
cubic interaction, which corresponds to Regime IV. In fact in Regime IV, we will repeat the above procedure.

Denote $a=\sqrt{C_0}$ and
 the ordinary differential equation \eqref{eq:har_bd_new} in $d$ dimensions can be solved analytically.  Denote
\begin{gather}\label{eq:fdr}
f_{a,d}(r)=
\begin{cases}
e^{ar}+e^{-ar},& \text{for } d=1,\\
I_0(ar),& \text{for } d=2,\\
(e^{ar}-e^{-ar})/r, &\text{for } d=3,
\end{cases}
\end{gather}
where $I_0(r)$ is the standard modified Bessel function $I_{\alpha}$ with $\alpha=0$.
Then the solution of Eq .\eqref{eq:har_bd_new} with prescribed Neumann boundary conditions reads as
\begin{equation}\label{eq:TF_har_pc}
n_0(r)=-\frac{\gm_0^2r^2}{2a^2}+\left(\frac{\tilde{\mu}}{a^2}-\frac{d\gm_0^2}{a^4}\right)+\frac{\gm_0^2R}{a^2f_{a,d}^{\prime}(R)}f_{a,d}(r).
\end{equation}
Inserting the above expression to the normalization condition that $\int_{\mathbb{R}^d}n_0(\bx)\,d\bx=1$, we
find chemical potential,
\begin{equation}\label{eq:TF_har_mu_pc}
\tmu=\frac{C_da^2}{R^d}+\frac{d\gm_0^2R^2}{2(d+2)}.
\end{equation}
Combining (\ref{eq:TF_har_mu_pc}) and (\ref{eq:TF_har_pc}), noticing the Dirichlet condition $n(R)=0$,
we have the equation for $R$,
\be
\left(\frac{(aR)^2}{d+2}-\frac{C_da^4}{\gm_0^2R^d}+d\right)f_{a,d}^{\prime}(R)=a^2Rf_{a,d}(R).
\ee

Thus, the free boundary $R$ can be calculated and $n_0(r)$ is then determined.

{\bf Regime III}, i.e. $\beta\ll\delta^{\frac{d+2}{d+4}}$, the $\beta$ term and the kinetic energy term are dropped, and the TF density profile is
\begin{equation}\label{ground_sol_har_ld}
n_{\rm TF}(r)=|\psi_{\rm TF}|^2=\frac{\gm_0^2(R^2-r^2)^2_+}{8(d+2)\delta},
\end{equation}
where $R=\left(\frac{(d+2)^2(d+4)C_d\delta}{\gm_0^2}\right)^{\frac{1}{d+4}}$. Again, the leading order approximations for chemical potential and energy,
 with the above TF densities, are $\mu_{\rm TF}=\frac{d}{2(d+2)}\left((d+2)^2(d+4)C_d\delta\gm_0^{d+2}\right)^{\frac{2}{d+4}}$ , $E_{\rm TF}=\frac{d+4}{d+6}\mu_{\rm TF}$ in $d$ dimensions.

{\bf Regime IV},  i.e. $\beta=-C_0\delta^{\frac{d+2}{d+4}}$ with $C_0>0$.
By a similar procedure as in Regime II, we'll get \eqref{eq:TFform} and	\be\label{eq:har_nbd_new}
\tmu=\frac{\gm_0^2r^2}{2}-{C_0}n_0-\partial_{rr}n_0(r)-\frac{d-1}{r}\partial_r n_0(r),
\ee
for $r\leq R$ and $n_0(s)=0$ for $s\ge R$, and  $n_0^\prime(0)=0$, $n_0(R)=0$, $n_0^{\prime}(R)=0$, where $R$ is the free boundary that has to be determined and $\tmu=\delta^{-\frac{2}{d+4}}\mu$.
 Again, let $a=\sqrt{C_0}$ and denote
\begin{gather}\label{eq:gddef}
g_{a,d}(r)=
\begin{cases}
\cos(ar),& \text{for } d=1,\\
J_0(ar),& \text{for } d=2,\\
\sin(ar)/r,& \text{for } d=3,
\end{cases}
\end{gather}
where $J_0(r)$ is the  Bessel function of the first kind $J_{\alpha}(r)$ with $\alpha=0$.
The solution of Eq. \eqref{eq:har_nbd_new} with the assigned Neumann boundary conditions can be written as:
\begin{equation}\label{eq:TF_har_nc}
n_0(r)=\frac{\gm_0^2r^2}{2a^2}-\left(\frac{\tmu}{a^2}+\frac{d\gm_0^2}{a^4}\right)-\frac{\gm_0^2{R}}{a^2g_{a,d}^{\prime}(R)}g_{a,d}(r).
\end{equation}
The chemical potential is then calculated from normalization condition as
\begin{equation}\label{eq:TF_har_mu_nc}
\tmu=-\frac{C_da^2}{R^d}+\frac{d\gm_0^2R^2}{2(d+2)}.
\end{equation}
Finally, the free boundary $R$ is  determined from the Dirichlet condition $n_0(R)=0$,
\be
\left(\frac{a^2R^2}{d+2}+\frac{C_da^4}{\gm_0^2{R^d}}-d\right)g_{a,d}^{\prime}(R)=a^2Rg_{a,d}(R).
\ee
After $R$ is computed, we then find  $n_0(r)$.

\begin{figure}
\includegraphics[width=.48\textwidth]{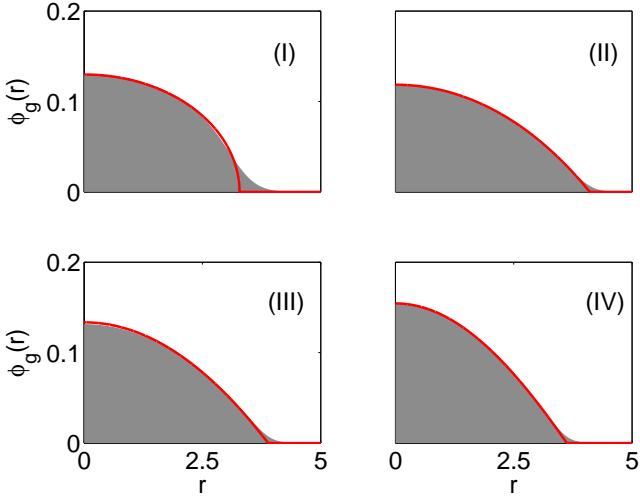}
\caption{Comparisons of 3D numerical ground states with TF densities, the harmonic potential case in region I, II, III and IV, which are define in Fig.~\ref{fig:regime}(a).
Red line: Thomas-Fermi approximation, and shaded area: numerical solution from the equation \eqref{eq:mgpe:d}. The parameters are chosen to be
$\gm=2$ and
(I) $\beta=1280$, $\delta=1$; (II) $\beta=828.7$, $\delta=1280$; (III) $\beta=1$, $\delta=1280$; (IV) $\beta=-828.7$, $\delta=1280$; respectively.}
\label{fig:tf_har}
\end{figure}

\begin{figure}
\centering
\begin{tabular}{c}
\subfloat[comparison of energy (harmonic potential case)]{\includegraphics[width=.48\textwidth]{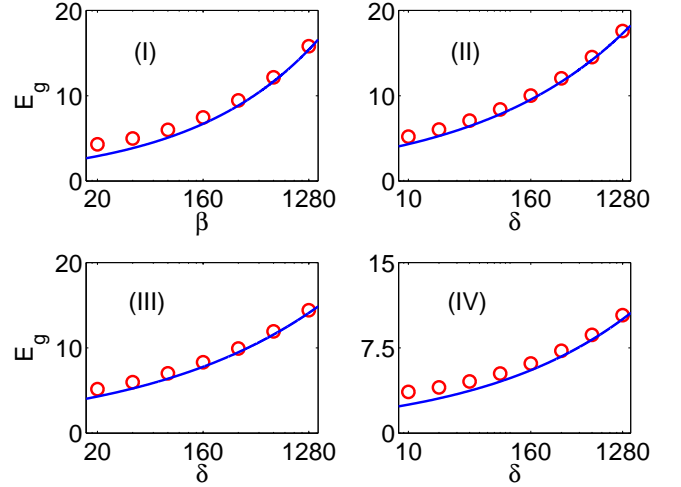}} \\
\subfloat[comparison of chemical potential (harmonic potential case)]{\includegraphics[width=.48\textwidth]{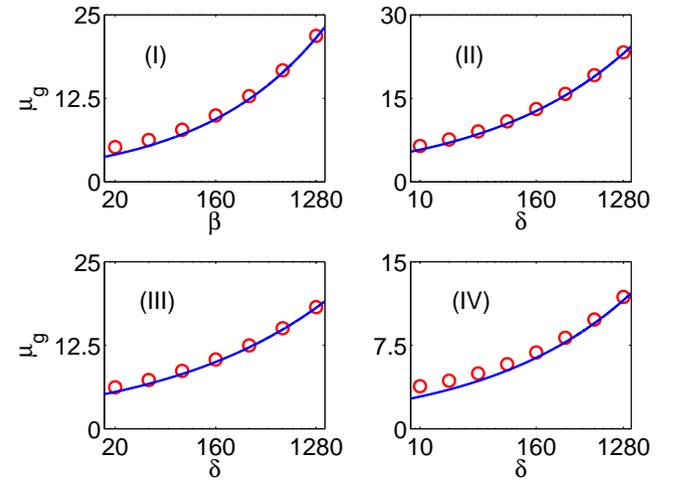}}
\end{tabular}
\caption{Comparisons of numerical energies and chemical potentials with TF approximations, the harmonic potential case. 3D problem is considered here.
 Blue line: Thomas-Fermi approximation, and red circles: numerical results obtained from the equation \eqref{eq:mgpe:d}. The parameters are chosen to be
$\gm=2$ and
(I) $\delta=1$, (II) $\beta=5\delta^{\frac{5}{7}}$, (III) $\beta=1$, (IV) $\beta=-5\delta^{\frac{5}{7}}$, respectively.}
\label{fig:E_har}
\end{figure}

In Fig.~\ref{fig:tf_har}, we compare the analytical TF densities \eqref{ground_sol_har_lb}, \eqref{ground_sol_har_ld} and \eqref{eq:TFform} with
the numerical results computed via full equation \eqref{eq:mgpe:d}
by the background Euler finite difference (BEFD) method \cite{BaoDu}.
We can observe  that in all the extreme regions, the analytical TF densities agree very well with
the full equation simulations. As a byproduct, we show the comparisons of the corresponding chemical potentials and energies in Fig.~\ref{fig:E_har}.

It has been shown that the usual TF densities provide accurate approximations for the density profiles for quasi-1D an 2D BECs. Indeed, we can check that for fixed three dimensional parameter $\beta$ and $\delta$, the effective contact interaction $\beta_1$ and HOI $\delta_1$ (or $\beta_2$ and $\delta_2$) for quasi-1D (2D) condensate, are
in the TF regime I, in the quasi-1D (2D) limit, i.e. $\gamma\to\infty$ ($\gamma\to0^+$). This justifies that the effective contact interactions is dominant for the dynamics in the quasi-1D
(2D) limit.

For instance, we know  $\beta_1\sim O(\gamma^{\frac{12}{7}})$ and $\delta_1\sim O(\gamma^{\frac{6}{7}})$ in quasi-1D limit, and it implies that
$\beta_1\gg \delta_1^{3/5}$ as $\gamma\gg1$. This immediately suggests that the TF density \eqref{ground_sol_har_lb} is a good approximation for the density profiles in quasi-1D limit regime,
which has been shown in Fig.~\ref{fig:cigar}. In the quasi-2D limit, i.e. $\gamma\to 0^+$, we find $\beta_{2}\gg \delta_2^{2/3}$ in view of $\beta_2\sim O(\gamma^{-3/2})$ and $\delta_2\sim O(\gamma^{-1/2})$, which again confirms that TF density {\eqref{ground_sol_har_lb} is a good approximations for the density profiles, as observed in Fig.~\ref{fig:disk}.


\subsection{\label{sec:box}TF approximation with box potential}
 In this section, we consider the  box potential case, which confines the BEC in a bounded domain $\{|\bx|\leq R\}$. Using similar method for the  harmonic potential case, we could obtain the analytical TF densities as the contact interaction  and/or HOI dominates the ground state in Eq. \eqref{eq:mgpe:d}.
In detail, we have the analytical TF densities for different regimes shown in Fig.~\ref{fig:regime}(b). Different from the harmonic potential case,
the borderline of the three regimes  is $\beta= O(\delta)$.


{\bf Regime I}, $\beta$ term is dominant, i.e. $\beta\gg1$ and $\delta=o(\beta)$. The kinetic term and the HOI term are dropped and the time independent MGPE equation in the radial variable $r$  becomes
\begin{equation}
\mu\psi(r)=\beta|\psi|^2\psi,\quad 0\le r=|\bx|< R,
\end{equation}
with boundary condition $\psi(R)=0$. Thus, the TF density is a constant, which can be uniquely determined by the normalization condition  $\|\psi\|=1$. Explicitly, TF density is given by $n_{\rm TF}(r)=|\psi|^2=\frac{C_d}{R^d}$, and  $\mu_{\rm TF}=\frac{C_d\beta}{R^d}$, where $C_d$ is defined in previous subsection.

It is obvious that the TF density is inconsistent with zero boundary condition, thus a boundary layer appears in the ground state density profiles \cite{Bao2007}.  In fact, as in \cite{Bao2007}, if $\delta\sim{o}(1)$, for $d=1$, to match the boundary layers at $x=\pm R$, an asymptotic analysis leads to the
following matched  density as $\beta\gg1$ for $0\le r=x\le R$,
\begin{equation}
n_{\rm as}(r)=|\psi_{\rm as}|^2=\frac{1}{2R}\left(\tanh(\sqrt{\mu_{\rm as}}(R-r))\right)^2,
\end{equation}
with the chemical potential $\mu_{\rm as}=\frac{1}{2R}\beta+\frac{1}{R}\sqrt{\frac{\beta}{2R}}$, and the energy $E_{\rm as}=\frac{1}{4R}\beta+\frac{2}{3R}\sqrt{\frac{\beta}{2R}}$.
For $d=2,3$, similar matched densities can be derived.

 From our numerical experience, the matched asymptotic density $n_{\rm as}$ provides much more accurate approximation to the ground state of Eq. \eqref{eq:mgpe:d}, than the TF density $n_{\rm TF}$,  in the parameter regimes $\beta\gg1$ and $\delta=O(1)$.

{\bf Regime II}, both $\beta$ and $\delta$ are important, i.e. $\beta=O(\delta)$ as $\delta\to\infty$. We assume that $\beta=C_0\delta$, with $\delta\gg1$ for some constant $C_0>0$.

Omitting the less important kinetic part, the radially symmetric time independent MGPE  reads
\be
\mu\psi(r)={C_0}\delta|\psi|^2\psi-\delta\nabla^2(|\psi|^2)\psi,\quad r<R,
\ee
with $\psi(R)=0$.
The above equation can be simplified for
density $n(r)=|\psi|^2$ in $d$ dimensions as
\be\label{eq:eqbox}
\frac{\mu}{\delta}={C_0}n(r)-\partial_{rr}n-\frac{d-1}{r}\partial_rn,
\ee
 with $n(R)=0$,  and at $r=0$ with $n^{\prime}(0)=0$.
Eq. \eqref{eq:eqbox} can be solved analytically.  Again, we introduce $a=\sqrt{C_0}$ and recall
function $f_{a,d}$ defined in \eqref{eq:fdr}.

The TF density, or solution of the boundary value problem \eqref{eq:eqbox}, is given explicitly as
\be
n_{\rm TF}(r)=|\psi_{\rm TF}|^2=\frac{\mu}{a^2\delta}\left[1-\frac{f_{a,d}(r)}{f_{a,d}(R)}\right],
\ee
with $\mu_{\rm TF}=C_da^2\delta/(R^d-d\frac{\int_0^Rf_{a,d}(r)r^{d-1}dr}{f_{a,d}(R)})$ 
and $E_{\rm TF}=\mu_{\rm TF}/2$, where $C_d$ is defined in Eq. \eqref{eq:cd}.

{\bf  Regime III}, $\delta$ term is dominant, i.e.  $\delta\gg1$, $\beta=o(\delta)$.  The kinetic  term and the $\beta$ term are dropped. The corresponding stationary MGPE
for the ground state reads
\be
\mu\psi=-\delta\nabla^2(|\psi|^2)\psi,
\ee
with boundary condition $\psi(R)=0$.

Solving the equation and using the normalization condition,  we obtain the TF density as
\be
n_{\rm TF}(r)=|\psi_{\rm TF}|^2=\frac{(d+2)C_d(R^2-r^2)}{2R^{d+2}},
\ee
with chemical potential $\mu_{\rm TF}={C_d}d(d+2)\delta/R^{d+2}$ and energy $ E_{\rm TF}=\mu_{\rm TF}/2$.

{\bf Regime IV}, i.e. $\beta=-C_0\delta$, with $\delta\gg1$ for some constant $C_0>0$.

 Intuitively, if $C_0$ is small, the repulsive
HOI $\delta$ term is dominant and the particle density will still occupy the entire domain; if $C_0$ is sufficiently large, the attractive $\beta$ interaction  becomes
the major effect, where the particles will be self trapped and the density profile will concentrate in a small portion of the domain.
Therefore, unlike the corresponding whole space case with harmonic potential,  we have two different situations here.

 By a similar procedure as in Regime II, we get
\be\label{eq:eqbox_nbd}
\frac{\mu}{\delta}=-{C_0}n(r)-\partial_{rr}n-\frac{d-1}{r}\partial_rn.
\ee
with $n(R^\prime)=0$ and $R^\prime$ to be determined.
In the first situation,  the density spreads over the whole domain and thus $R^\prime=R$;
in the second situation, the density would concentrate and $0<R^\prime<R$.

{\it Case I}, i.e. $C_0\leq C_{\rm cr}$, where $C_{\rm cr}=\hat{R}^2/R^2$ and $\hat{R}$ is  the
first positive root of $g_{a,d}^\prime(r/a)=0$  defined in Eq. \eqref{eq:gddef} with $a=\sqrt{C_0}$\,.
As mentioned before, because of the relatively weak attractive interaction, we still have the following boundary conditions at the boundary:
$n(R)=0$, $n^{\prime}(0)=0$.

The TF density, or solution of Eq. \eqref{eq:eqbox_nbd}, can be expressed as:
\be
n_{\rm TF}=|\psi_{\rm TF}|^2=-\frac{\mu}{a^2\delta}\left[1-\frac{g_{a,d}(r)}{g_{a,d}(R)}\right],
\ee
with $\mu_{\rm TF}=C_da^2\delta/(d\frac{\int_0^Rg_{a,d}(r)r^{d-1}dr}{g_{a,d}(R)}-R^d)
$ and
$E_{\rm TF}=\mu_{\rm TF}/2$, where $C_d$ is given in \eqref{eq:cd}.

If $aR>\hat{R}$, we know from the properties of $g_{a,d}(r)$ that $g_{a,d}(r)$ ($r\in[0,\hat{R}]$) would take any value between the maximum (positive)  and minimum (negative) of
$g_{a,d}(r)$ ($r\ge0$). Then $1-g_{a,d}(r)/g_{a,d}(R)$ would change sign for $r\in[0,\hat{R}]$, when $g_{a,d}(r)$ takes value around $r_0\in(0,\hat{R})$ such that $g_{a,d}(r_0)=g_{a,d}(R)$.
On the other hand, since the density must be nonnegative, $1-g_{a,d}(r)/g_{a,d}(R)$ can not change sign in $[0,R]$. So we conclude that $aR\leq\hat{R}$, i.e. the condition
$C_0\leq C_{\rm cr}$ is necessary.

$g_{a,d}^\prime$ at $r/a$ can be computed as
\begin{gather}\label{def:box_nbd_df}
g_{a,d}^{\prime}(r/a)=
\begin{cases}
-a\sin(r),&  d=1,\\
-aJ_1(r),&  d=2,\\
a^2(r\cos(r)-\sin(r))/r^2,&  d=3,
\end{cases}
\end{gather}
and we have
for 1D case, $\hat{R}=\pi$;
for 2D case, $\hat{R}=3.8317\cdots$;
for 3D case, $\hat{R}=4.4934\cdots$. \\

 {\it Case II}, $C_0>C_{\rm cr}$.
As observed above, the density profiles may be away from the boundaries of the domain. Thus,  free boundary conditions should be used as
$n(\tilde{R})=0$, $n^{\prime}(\tilde{R})=0$, $n^{\prime}(0)=0$, where $\tilde{R}<R$ is the boundary for the TF density that we want to find.

   Hence domain $[0,\tilde{R}]$ replaces the domain $[0,R]$ in {\it Case I},  and we have extra boundary condition $n^{\prime}(\tilde{R})=0$.
   Denoting $a=\sqrt{C_0}$ and using the solution in {\it Case I}, we get $g^\prime_{a,d}(\tilde{R})=0$, and  $a\tilde{R}\leq \hat{R}$, where
   both conditions can only be satisfied if and only if $a\tilde{R}=\hat{R}$. So, we identify that $\tilde{R}=\hat{R}/a<R$.

\begin{figure}
\includegraphics[width=.48\textwidth]{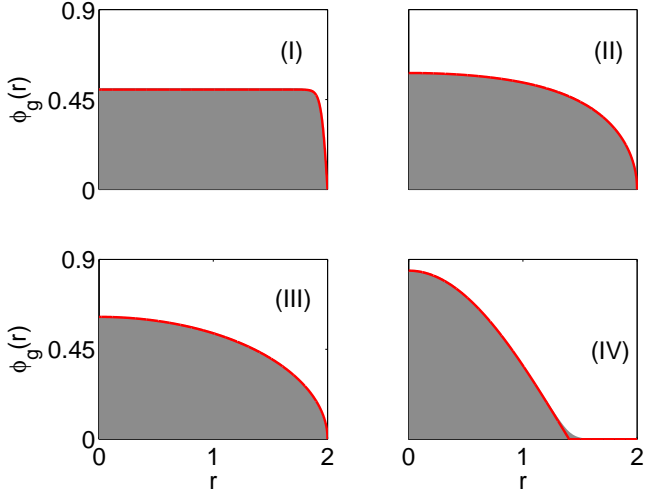}
\caption{Comparisons of 1D numerical ground states with TF densities, the box potential case in region I, II, III and IV, which are define in Fig.~\ref{fig:regime}(b). Red line: analytical TF approximation,  and shaded area: numerical solution obtained from \eqref{eq:mgpe:d}.
Domain is $\{r|0\le r<2\}$ and the corresponding $\beta$'s and $\delta$'s are (I) $\beta=1280$, $\delta=1$; (II) $\beta=320$, $\delta=160$; (III) $\beta=1$, $\delta=160$; (IV) $\beta=-400$, $\delta=80$.}
\label{fig:tf_box}
\end{figure}

\begin{figure}
\centering
\begin{tabular}{c}
\subfloat[comparison of energy]{\includegraphics[width=.48\textwidth]{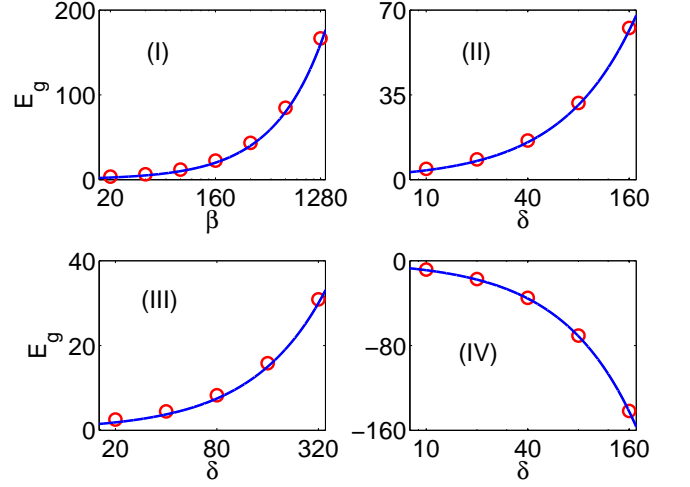}} \\
\subfloat[comparison of chemical potential]{\includegraphics[width=.48\textwidth]{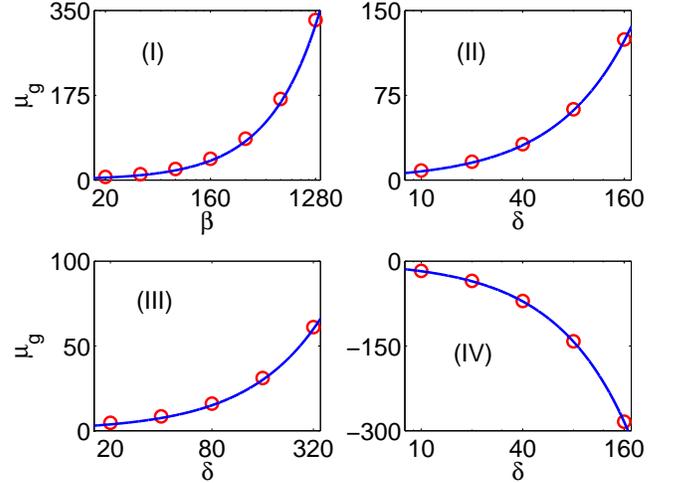}}
\end{tabular}
\caption{Comparisons of numerical energies and chemical potentials with TF approximations, the box potential case. 1D problem is considered here. Blue line: analytical TF approximation, and red
circles: numerical results obtained from \eqref{eq:mgpe:d}.
The parameters are chosen to be (I) $\delta=1$, (II) $\beta=2\delta$, (III) $\beta=1$, (IV) $\beta=-5\delta$, respectively, and domain is $\{r|0\le r<2\}$.}
\label{fig:E_box}
\end{figure}

Replacing  $R$ with $\hat{R}/a$ in TF solution of {\it Case I}, we obtain the analytical TF density as
\be
n_{\rm TF}(r)=|\psi_{\rm TF}|^2=\frac{C_da^d}{\hat{R}^d}\left[1-\frac{g_{a,d}(r)}{g_{a,d}(\frac{\hat{R}}{a})}\right],
\ee
with $\mu_{\rm TF}=-C_da^{d+2}\delta/\hat{R}^d$ and $E_{\rm TF}=\mu_{\rm TF}/2$,
where $\hat{R}$ is defined in {\it Case I}.

We compare in Fig.~\ref{fig:tf_box} the analytical TF densities listed above with the ground state obtained from numerical results via Eq. \eqref{eq:mgpe:d} computed by the BEFD method \cite{BaoDu} in various
parameter regimes discussed above.
Fig.~\ref{fig:tf_box} shows our analytical TF densities are very good approximations for the ground states. We also make comparisons for chemical potentials and energies  between the TF approximations and the numerical values by solving Eq. \eqref{eq:mgpe:d}  in Fig.~\ref{fig:E_box}.

\section{\label{sec:conclusion}conclusion}
We have presented the mean-field modified Gross-Pitaevskii equations for quasi-1D, Eq.~\eqref{3to1model}, and quasi-2D, Eq.~\eqref{3to2model}, BECs with   higher-order interaction (HOI) term.
These equations are based on a rigorous
dimension reduction from the full 3D MGPE with the assumptions that the  energy separations
in  radial and longitudinal directions scales differently in the strongly anisotropic aspect ratio limit,
and the wave function can be separated into radial and longitudinal variables.  By carefully studying
the energy separation, we obtain the correct  radial or longitudinal states used in the dimension reduction.  In particular, it is quite interesting
that the radial states has to be taken in the form different from the ground state of radial harmonic potential in the quasi-1D BEC, which
is counterintuitive compared with the conventional GPE.
Our result shows that quasi-1D and quasi-2D
BECs with HOI are governed by a modified contact interaction term
and a modified HOI term, and all the equations for quasi-1D and quasi-2D BECs have the same form as the 3D MGPE.

We have computed the ground states of our 1D and 2D equations
numerically and compared them with the ground states of the 3D MGPE, and we find excellent agreements.
We have also completely  determined   Thomas-Fermi approximation  in various parameter regimes with both box potential and harmonic potential, for the 1D, 2D and 3D cases.
In presence of HOI, TF approximations become very complicated as HOI competes with contact interaction.


\appendix
\section{Derivation of the quasi-1D equation}\label{appendix:3to1}
Under the assumption in Sec.~\ref{sec:3to1},
we take the ansatz
\begin{equation}\label{factorization}
\psi(x,y,z,t)=e^{-i\mu_{2D}t}\omega_{2D}(x,y)\psi_{1D}(z,t),
\end{equation}
where the transverse state is frozen, i.e. $\omega_{2D}$ is the radial minimum energy state  and the energy separation is much larger in the radial direction than the longitudinal $z$ direction.

Substitute \eqref{factorization} into Eq. \eqref{eq:mgpe}, we can get the equations for $\psi_{1D}$  for appropriate $\mu_{2D}$ as
\begin{widetext}
\be
i\p_t\psi_{1D}(z,t)=\left[-\frac{1}{2}\partial_{zz}+V_{1D}(z)+\beta_1|\psi_{1D}|^2-
\delta_1(\partial_{zz}|\psi_{1D}|^2)\right]\psi_{1D},\label{zmodel}
\ee
\end{widetext}
where $V_{1D}(z)=\frac12z^2$,
\begin{subequations}\label{para:bd}
\begin{align}
\beta_1&=\beta\iint|\omega_{2D}|^4dxdy+\delta\iint|\nabla_{\perp}|\omega_{2D}|^2|^2dxdy,\label{beta_z}\\
\delta_1&=\delta\iint|\omega_{2D}|^4dxdy,\label{delta_z}
\end{align}
\end{subequations}
and $\nabla_{\perp}=(\p_x,\p_y)^T$.
It remains to determine $\omega_{2D}$ and we are going to use the criteria that the energy separations scale differently in different directions. In order to
do this, we need calculate the energy scale in $z$ direction. Hence, we take the stationary states (ground states) of \eqref{zmodel} as
\be
\psi_{1D}(z,t)=e^{-i\mu_{1D}t}\phi_{1D}(z)\label{eq:psi1d}.
\ee
Combining Eqs. \eqref{factorization} and \eqref{eq:psi1d}, following the way to find Eq. \eqref{zmodel},
we can derive the equations for $\omega_{2D}(x,y)$ as
\begin{widetext}
\begin{equation}
\mu_{2D}\omega_{2D}=-\frac{1}{2}\nabla^2_{\perp}\omega_{2D}+V_{2D}(r)\omega_{2D}+\beta_2|\omega_{2D}|^2\omega_{2D}-\delta_2(\nabla^2_{\perp}|\omega_{2D}|^2)\omega_{2D},\label{rmodel}
\end{equation}
\end{widetext}
where $\nabla^2_\perp=\p_{xx}+\p_{yy}$, the radially symmetric potential $V_{2D}=\frac{\gamma^2}{2}(x^2+y^2)$,
\begin{subequations}\label{para:bd2}
\begin{align}
\beta_2&=\beta\int|\phi_{1D}|^4dz+\delta\int|\partial_{z}|\phi_{1D}|^2|^2dz,\label{beta_r}\\
\delta_2&=\delta\int|\phi_{1D}|^4dz.\label{delta_r}
\end{align}
\end{subequations}

To determine  the frozen state  $\omega_{2D}$, we need minimize the energy of Eq. \eqref{rmodel}, while parameters $\beta_2$ and $\delta_2$ depends on $\phi_{1D}$. So actually, we need solve a coupled system together for $\omega_{2D}$ and $\phi_{1D}$. To this purpose, we will consider the problem  in the quasi-1D limit $\gamma\to\infty$.  Intuitively, transverse direction is almost compressed to a Dirac function as $\gamma\to\infty$, so that a proper scaling is needed to obtain the correct form of $\omega_{2D}$.

We will determine $\omega_{2D}$ via a self consistent iteration as follows:
given some $\beta_2$ and $\delta_2$, under proper scaling as $\gamma\to\infty$,  (i)
drop the less important part to get approximate $\omega_{2D}$, (ii)
 put $\omega_{2D}$ into Eq. \eqref{zmodel} to determine the longitudinal ground state $\phi_{1D}$,
(iii) use $\phi_{1D}$ to compute $\beta_2$ and $\delta_2$, and then (iv) check if it is consistent.

In the quasi-1D regime, $\gamma\to\infty$, similarly to the conventional GPE case, due to the strong confinement in transverse direction, the ground state solution $\phi_{1D}$ is very flat in $z$ direction, as both nonlinear terms exhibit repulsive interactions. It is easy to get the scalings of $\int|\partial_z|\phi_{1D}|^2|^2dz=O(L^{-3})$, $\int|\phi_{1D}|^4dz=O(L^{-1})$, where $L$ indicates the correct length scale of $\phi_{1D}$.  Therefore $\beta_2$ and $\delta_2$ are of the same order by definition, since $L\to \infty$ in the quasi-1D limit.

For mathematical convenience,   we introduce $\vep=1/\sqrt{\gm}$ such that $\vep\to0^+$. In the radial variable, introduce the new scale $\tr=r/\vep^{\alpha}$ and $\tilde{\omega}(\tr)=\vep^{\alpha}\omega_{2D}(r)$ such that $\tr\sim O(1)$ and $\|\tilde{\omega}\|=1$, then \eqref{rmodel} becomes
\begin{equation}\label{eq:r_eig_eq_m}
\mu_{2D}\tilde{\omega}=-\frac{\nabla^2_{\perp}\tilde{\omega}}{2\vep^{2\alpha}}
+\frac{\tr^2\tilde{\omega}}{2\vep^{4-2\alpha}}+\frac{\beta_2}{\vep^{2\alpha}}\tilde{\omega}^3-\frac{\delta_2}{\vep^{4\alpha}}\nabla^2_{\perp}(|\tilde{\omega}|^2)\tilde{\omega}.
\end{equation}

 Noticing  that the term $\beta_2/\vep^{2\alpha}\tilde{\omega}^3$ can  be always neglected compared to the last term since $\beta_2\sim\delta_2$ and $\vep^{-\alpha}\ll\vep^{-3\alpha}$ as $\vep\to0^{+}$. On the other hand, $\beta_2$ and $\delta_2$ are both repulsive interactions while only the potential term confines the condensate. Thus, the correct  leading effects (HOI or kinetic term)
 should be balanced with the potential term. Now, we are only left with two possibilities:

 {\it Case I}, $-\frac{1}{2\vep^{2\alpha}}\tilde{\nabla^2}_{\perp}\tilde{\omega}$ is balanced with term $\frac{\tr^2}{2\vep^{4-2\alpha}}\tilde{\omega}$, and $\frac{\delta_2}{\vep^{4\alpha}}\tilde{\nabla^2}_{\perp}(|\tilde{\omega}|^2)\tpsi$ is smaller. In this case, $\vep^{2\alpha}\sim\vep^{4-2\alpha}$. So we get $\alpha=1$. Besides, we also need $\vep^{-2\alpha}\gg\frac{\delta_2}{\vep^{4\alpha}}$, i.e. $\delta_2\ll\vep^2$.

 {\it Case II}, $\frac{\delta_2}{\vep^{4\alpha}}\tilde{\nabla^2}_{\perp}(|\tilde{\omega}|^2)\tilde{\omega}$ is balanced with term $\frac{\tr^2}{2\vep^{4-2\alpha}}\tilde{\omega}$, and $-\frac{1}{2\vep^{2\alpha}}\tilde{\nabla^2}_{\perp}\tilde{\omega}$ is much smaller. In this case, $\frac{\delta_2}{\vep^{4\alpha}}\sim\frac{1}{\vep^{4-2\alpha}}$ and $\vep^{-2\alpha}\ll\frac{1}{\vep^{4-2\alpha}}$, i.e. $\alpha<1$ and $\delta_2\sim\vep^{6\alpha-4}$.

We will check if the scaling is consistent for each case.

\,

{\it Case I}. Since $\alpha=1$ , we have $\omega_{2D}$ as the ground state of radial harmonic
oscillator,
 \begin{equation}\label{r_sol1}
\omega_{2D}(r)=\frac{1}{\sqrt{\pi\vep^2}}e^{-\frac{r^2}{2\vep^2}},
\end{equation}
and
\begin{equation*}
\iint|\omega_{2D}|^4dxdy=\frac{1}{2\pi\vep^2},\quad
\iint|\nabla_{\perp}(|\omega_{2D}|^2)|^2dxdy=\frac{1}{\pi\vep^4}.
\end{equation*}
Recalling $\beta_1$ and $\delta_1$ in Eq. \eqref{para:bd}, the parameters are in TF regime I (cf.  Sec.~\ref{sec:3to1}), so in $z$ direction  we can get the approximate solution from Sec.~\ref{sec:3to1} as:
\begin{equation}\label{z_sol}
\phi_{1D}\approx\sqrt{\frac{\left((z^{*})^2-z^2\right)_+}{2\beta_1}},\quad z^*=\left(\frac{3\beta_1}{2}\right)^{\frac{1}{3}},
\end{equation}
By definition of $\delta_2$ \eqref{delta_r}, we obtain
\begin{equation}\label{db1_3to1}
\delta_2=\delta\int|\phi_{1D}|^4dz=\frac{3\delta}{5}\left(\frac{2}{3\beta_1}\right)^{\frac{1}{3}},
\end{equation}
while
\begin{equation}\label{bd1_3to1}
\beta_1\sim\delta\iint|\nabla_{\perp}(|\omega_{2D}|^2)|^2dxdy=\frac{\delta}{\pi\vep^4}.
\end{equation}
Combining \eqref{db1_3to1}) and \eqref{bd1_3to1}, we get $\delta_2=O(\vep^{\frac{4}{3}})$. But it contradicts with the requirement that $\delta_2\ll\vep^2$.
Thus {\it Case I} is inconsistent.

{\it Case II}. As $\delta_2$ term is more significant than the kinetic term, we solve  $\mu_{2D}=r^2/2\vep^4-\delta_2\nabla_{\perp}^2|\omega_{2D}|^2$
within the support of $\omega_{2D}(r)$ and get
 \begin{equation}\label{r_sol2}
\omega_{2D}(r)=\frac{(R^2-r^2)_+}{\sqrt{32\vep^4\delta_2}},\, R=2a\vep,\,a=\left(\frac{3\delta_2}{2\pi\vep^2}\right)^{\frac{1}{6}}.
\end{equation}
Hence, we know
\be
\iint|\omega_{2D}|^4dxdy=\frac{3}{10\delta_2}\left(\frac{3\delta_2}{2\pi\vep^2}\right)^{\frac{2}{3}},
\ee
 \be\label{r_order2}
 \iint|\nabla_{\perp}|\omega_{2D}|^2|^2dxdy=\frac{1}{2\delta_2\vep^2}\left(\frac{3\delta_2}{2\pi\vep^2}\right)^{\frac{1}{3}}.
\ee
Again, recalling $\beta_1$ and $\delta_1$ in Eq. \eqref{para:bd}, the parameters are in TF regime I (cf.  Sec.~\ref{sec:3to1}), so in $z$ direction  we can get the approximate solution from Sec.~\ref{sec:3to1} as Eq. \eqref{z_sol}. Having $\phi_{1D}(z)$ Eq. \eqref{z_sol}, we can compute
\begin{equation}\label{db2_3to1}
\delta_2=\delta\int|\phi_{1D}|^4dz=\frac{3\delta}{5}\left(\frac{2}{3\beta_1}\right)^{\frac{1}{3}},
\end{equation}
while
\begin{equation}\label{bd2_3to1}
\beta_1\sim\delta\int|\nabla_{\perp}|\omega_{2D}|^2|^2dxdy=\left(\frac{3}{2\pi}
\right)^{\frac{1}{3}}\frac{\delta}{2\vep^{\frac{8}{3}}\delta_2^{\frac{2}{3}}}.
\end{equation}
Combining \eqref{db2_3to1} and \eqref{bd2_3to1}, we find $\delta_2=\frac{2\cdot 3^{\frac{5}{7}}\pi^{\frac{1}{7}}\delta^{\frac{6}{7}}\vep^{\frac{8}{7}}}{5^{\frac{9}{7}}}$,
$\beta_1\sim\frac{5^{\frac{6}{7}}}{3^{\frac{1}{7}}\cdot4\pi^{\frac{3}{7}}}\delta^{\frac{3}{7}}\gm^{\frac{12}{7}}$.
Noticing the requirement that $\delta_2\sim\vep^{6\alpha-4}$, we get $\alpha=6/7$, and it satisfies the other constraint $\alpha<1$. Thus, {\it Case II} is self consistent.
Therefore, for the quasi-1D BEC, this is the case that we should choose to derive the mean field equation and $\beta_1$, $\delta_1$ can be obtained as in Eq. \eqref{eq:1dpara}.

To summarize, we identify that $\omega_{2D}$ should be taken as Eq. \eqref{r_sol2} and the mean field equation Eq. \eqref{3to1model} for quasi-1D BEC is derived.

With this explicit form of the approximate solutions, we can further get the leading order approximations for chemical potential and energy for the original 3D problem. It turns out that $\mu_g^{3D}\approx\frac{9}{8}\mu_{2D} \text{ and } E_g^{3D}\approx\frac{7}{8}\mu_{2D},$ where $\mu_{2D}$ is computed approximately as before.

\section{Derivation of the quasi-2D equation}\label{appendix:3to2}
Under the assumption in Sec.~\ref{sec:3to1},
we take the ansatz
\begin{equation}\label{factorization2}
\psi(x,y,z,t)=e^{-i\mu_{1D}t}\omega_{1D}(z)\psi_{2D}(x,y,t),
\end{equation}
where the longitudinal state is frozen, i.e. $\omega_{1D}$ is the minimum energy state  and the energy separation is much larger in the  longitudinal $z$ direction than the radial direction.

Plugging Eq.~\eqref{factorization2} into Eq. \eqref{eq:mgpe}, we can get the equations for $\psi_{2D}$  with appropriate $\mu_{1D}$ as
\begin{widetext}
\be
i\p_t\psi_{2D}(x,y,t)=\left[-\frac{1}{2}\nabla^2_\perp+V_{2D}(x,y)+
\beta_2|\psi_{1D}|^2-\delta_2(\nabla^2_{\perp}|\psi_{2D}|^2)\right]\psi_{2D},\label{rmodel:2}
\ee
\end{widetext}
where the radially symmetric potential $V_{2D}(r)=\frac12r^2$ and
\begin{subequations}\label{para:bd:2d}
\begin{align}
\beta_2&=\beta\int|\omega_{1D}|^4dxdy+\delta\int|\partial_{z}|\omega_{1D}|^2|^2dz,\label{beta_2:r}\\
\delta_2&=\delta\int|\omega_{1D}|^4dxdy,\label{delta_2:r}
\end{align}
\end{subequations}
with $\nabla_{\perp}=(\p_x,\p_y)^T$ and $\nabla^2_\perp=\partial_{xx}+\partial_{yy}$.
It remains to determine $\omega_{1D}$ and we are going to use the same idea as that in the quasi-1D BEC. In order to
do this, we need calculate the energy scale in $r$ direction. Hence, we take the stationary states (ground states) of Eq.~\eqref{rmodel:2} as
\be
\psi_{2D}(r,t)=e^{-i\mu_{2D}t}\phi_{2D}(r).\label{eq:psi2d}
\ee
Combining Eq. \eqref{factorization2} with Eq. \eqref{eq:psi2d},
we can derive the equations for $\omega_{1D}(z)$ as
\begin{widetext}
\begin{equation}
\mu_{1D}\omega_{1D}=-\frac{1}{2}\partial_{zz}\omega_{1D}+V_{1D}(z)\omega_{1D}+\beta_1|\omega_{1D}|^2\omega_{1D}-\delta_1(\partial_{zz}|\omega_{1D}|^2)\omega_{1D},\label{rmodel:2d}
\end{equation}
\end{widetext}
where $V_{1D}(z)=\frac{z^2}{2\gamma^2}$,
\begin{subequations}\label{para:bd2:2d}
\begin{align}
\beta_1&=\beta\int|\phi_{2D}|^4dz+\delta\int|\nabla_{\perp}|\phi_{2D}|^2|^2dz,\label{beta_r:2d}\\
\delta_1&=\delta\int|\phi_{2D}|^4dz.\label{delta_r:2d}
\end{align}
\end{subequations}

We proceed similarly to the quasi-1D case. For mathematical convenience, denote $\vep=\sqrt{\gm}$ such that $\vep\rightarrow{0^+}$.
 Rescale $z$ variable as $\tilde{z}=z/\vep^{\alpha}$, $\tilde{\omega}(\tilde{z})=\vep^{\frac{\alpha}{2}}\omega_{1D}(z)$ for some $\alpha>0$.
 By removing the tildes,  Eq. \eqref{rmodel:2d} becomes
\begin{equation}\label{eq:z_eig_eq_m}
\mu_{1D}\omega=-\frac{1}{2\vep^{2\alpha}}\partial_{zz}\omega+\frac{z^2}{2\vep^{4-2\alpha}}\omega
+\frac{\beta_1}{\vep^{\alpha}}\omega^3-\frac{\delta_1}{\vep^{3\alpha}}(\partial_{zz}|\omega|^2)\omega.
\end{equation}
Assuming that the scale is correct, then  $\omega$ will be a regular function, independent of $\vep$ so that its norm will be $O(1)$.
Now, we will determine the scale similarly to the quasi-1D BEC. Intuitively, by the same reason in the quasi-1D case, the term $\frac{\beta_1}{\vep^{\alpha}}\omega^3$ can always be neglected compared to the HOI term.  In addition, potential term is the only effects that confine the condensate, which can not be neglected. Then, there are two possibilities:

 {\it Case I}. $-\frac{1}{2\vep^{2\alpha}}\partial_{zz}\omega$ is balanced with term $\frac{z^2}{2\vep^{4-2\alpha}}\omega$, and $\frac{\delta_1}{\vep^{3\alpha}}(\partial_{zz}|\omega|^2)\omega$ is much smaller. In this case, $\vep^{2\alpha}\sim\vep^{4-2\alpha}$. So we get $\alpha=1$. Besides, we also need $\vep^{-2\alpha}\gg\frac{\delta_1}{\vep^{3\alpha}}$, i.e. $\delta_1\ll\vep$.

{\it Case II}.  $\frac{\delta_1}{\vep^{3\alpha}}(\partial_{zz}|\omega|^2)\omega$ is balanced with term $\frac{z^2}{2\vep^{4-2\alpha}}\omega$, and $-\frac{1}{2\vep^{2\alpha}}\partial_{zz}\omega$ is much smaller. In this case, $\frac{\delta_1}{\vep^{3\alpha}}\sim\frac{1}{\vep^{4-2\alpha}}$ and $\vep^{-2\alpha}\ll\frac{1}{\vep^{4-2\alpha}}$, i.e. $\alpha<1$ and $\delta_1\sim\vep^{5\alpha-4}$.

Now, we check the consistency of each case.

{\it Case I}. Since $\alpha=1$, we can obtain $\omega_{1D}(z)$ as the ground state
of longitudinal harmonic oscillator as
 \begin{equation}\label{z_sol1}
\omega_{1D}(z)=\left(\frac{1}{\pi\vep^2}\right)^{\frac{1}{4}}e^{-\frac{z^2}{2\vep^2}},
\end{equation}
and the following quantities can be calculated:
\begin{equation}\label{z_order1}
\int|\omega_{1D}|^4dz=\frac{1}{\sqrt{2\pi}\vep} \text{ , } \int|(|\omega_{1D}|^2)^{\prime}|^2dz=\frac{1}{\sqrt{2\pi}\vep^3}.
\end{equation}
By examining $\beta_2$ and $\delta_2$ in Eq. \eqref{rmodel:2}, we find $\beta_2$ is dominant as $\vep\to0^+$ and the ground state
$\phi_{2D}(r)$ can be obtained as TF approximation in the parameter regime I as shown in Sec.~\ref{sec:har},
\begin{equation}\label{r_sol}
\phi_{2D}(r)=\sqrt{\frac{(R^2-r^2)_+}{2\beta_2}},\text{ where } R=\left(\frac{4\beta_2}{\pi}\right)^{\frac{1}{4}}.
\end{equation}
Then we can compute
\begin{equation}\label{r1_order}
\iint|\phi_{2D}|^4dxdy=\frac{2}{3\sqrt{\pi\beta_2}},
\end{equation}
\begin{equation}\label{r2_order}
\iint|\nabla_{\perp}(|\phi_{2D}|^2)|^2dxdy=\frac{2}{\beta_2}.
\end{equation}
Having $\phi_{2D}$, we can check the consistency of {\it Case I}.
By definition of $\delta_1$ in Eq.~\eqref{para:bd2:2d}, we get
\begin{equation}\label{db1_3to2}
\delta_1=\delta\iint|\phi_{2D}|^4dxdy=\frac{2\delta}{3\sqrt{\pi\beta_2}},
\end{equation}
while it follows from the definition of $\beta_2$ in Eq. \eqref{para:bd:2d},
\begin{equation}\label{bd1_3to2}
\beta_2\sim\delta\int|(|\omega_{1D}|^2)^{\prime}|^2dz=\frac{\delta}{\sqrt{2\pi}\vep^3}.
\end{equation}
Combining Eqs.~\eqref{db1_3to2} and \eqref{bd1_3to2}, we obtain
$\delta_1=\frac{2}{3}\sqrt{\frac{\delta}{\pi}}\left(2\pi\right)^{\frac{1}{4}}
\vep^{\frac{3}{2}}=O(\vep^{\frac{3}{2}})=
o(\vep)$, which satisfies the requirement for $\delta_1$.
Thus, {\it Case I} is self consistent.

{\it Case II}. In this case, we solve equation $\mu_{1D}=\frac{z^2}{2\vep^4}-\delta_1\partial_{zz}|\omega_{1D}|^2$ within the support
of $\omega_{1D}$ and get
 \begin{equation*}
\omega_{1D}(z)=\frac{\left((z^{*})^2-z^2\right)_+}{2\vep^2\sqrt{6\delta_1}},\quad z^{*}=\left(\frac{45\delta_1\vep^4}{2}\right)^{\frac{1}{5}}.
\end{equation*}
Then we have the identities as
\begin{equation}\label{z_order2}
\int|\omega_{1D}|^4dz=\frac{2}{63}\left(\frac{45}{2}\right)^{\frac{4}{5}}\left(\vep^4\delta_1\right)^{-\frac{1}{5}},
 \end{equation}
\begin{equation}
 \int|(|\omega_{1D}|^2)^{\prime}|^2dz=\frac{2}{21}\left(\frac{45}{2}\right)^{\frac{2}{5}}\left(\vep^4\delta_1\right)^{-\frac{3}{5}}.
\end{equation}
In the quasi-2D limit regime, i.e. $0<\vep\ll1$,  by the definitions of $\beta_2$ and $\delta_2$ in Eq.~\eqref{para:bd:2d}, we find $\beta_2$
is dominant and $\phi_{2D}$ can be obtained as the TF density in parameter regime I shown in Sec.~\ref{sec:har}, which is exactly the same
as Eq.~\eqref{r_sol}.

Similarly to the previous case, we can calculate
\begin{equation}\label{db2_3to2}
\delta_1=\delta\iint|\phi_{2D}|^4dxdy=\frac{2\delta}{3\sqrt{\pi\beta_2}},
\end{equation}
where
\begin{equation}\label{bd2_3to2}
\beta_2\sim\delta\int|(|\omega_{1D}|^2)^{\prime}|^2dz=\frac{2\delta}{21}\left(\frac{45}{2}\right)^{\frac{2}{5}}\left(\vep^4\delta_1\right)^{-\frac{3}{5}}.
\end{equation}
Combining Eqs.~\eqref{db2_3to2} and \eqref{bd2_3to2}, we can get $\delta_1\approx\frac{2}{45}\left(\frac{105\delta}{\pi}\right)^{\frac{5}{7}}\vep^{\frac{12}{7}}$.
But the requirement is $\delta_1\sim\vep^{5\alpha-4}$ and we get $\alpha=8/7$. This contradicts with the other requirement that $\alpha<1$.
In other words, {\it Case II} is inconsistent.

In summary, {\it Case I} is true and $\omega_{1D}$ should be chosen as Eq.~\eqref{z_sol1}. Thus, mean-field equation for quasi-2D BEC is derived in Eq.~\eqref{3to2model}
with given constants in Eq.~\eqref{beta_2}.

\bigskip

\section{Rescaling with harmonic potential}\label{appendix:TF}
 In this section, we show how to distinguish  the four extreme regions in the TF approximations for Eq.~\eqref{eq:mgpe:d}.

 In $d$ ($d=3,2,1$) dimensions, introduce $\tilde{\bx}=\frac{\bx}{x_s}$, and
$\tilde \psi(\tilde{\bx})=x_s^{d/2}\psi(\bx)$ such that $x_s$ is the Thomas-Fermi radius of
the wave function and then the Thomas-Fermi radius in the new scaling is at $O(1)$. It's easy to check that such scaling conserves the normalization condition Eq.~\eqref{eq:normalization}. Substituting
$\tilde{\bx}$ and $\tilde\psi$ into the time-independent version of \eqref{eq:mgpe:d} and then removing
all $\tilde{\ }$, we get
\begin{equation*}
\frac{\mu}{x_s^2}\psi=-\frac{1}{2x_s^4}\nabla^2\psi+\frac{\gm_0^2|\bx|^2}{2}\psi
+\frac{\beta}{x_s^{2+d}}\psi^3-\frac{\delta}{x_s^{4+d}}\nabla^2(|\psi|^2)\psi.
\end{equation*}
Since it is assumed $x_s$ is the length scale and the potential term would be $O(1)$. To balance  the confinement with repulsive interactions, we need $\frac{\beta}{x_s^{2+d}}\sim O(1)$ and/or $\frac{\delta}{x_s^{4+d}}\sim O(1)$. For simplicity, we require $\frac{\delta}{x_s^{4+d}}=1$, then $x_s=\delta^{\frac{1}{4+d}}$, and further $\beta\sim O(x_s^{2+d})\sim O(\delta^{\frac{2+d}{4+d}})$. So the borderline case is
$\beta=C_0\delta^{\frac{2+d}{4+d}}$. If $C_0\gg1$, $\beta$ term is much more significant than the $\delta$ term; if $|C_0|\ll 1$,  $\delta$ term is much more significant than the $\beta$ term.


\end{document}